\documentclass[12pt,amsmath,amssymb,]{revtex4}
\usepackage{amsfonts}
\usepackage{amsmath}
\usepackage{amssymb}
\usepackage{array}
\usepackage{bm}
\usepackage{braket}
\usepackage{breqn}
\usepackage{color}
\usepackage{dcolumn}
\usepackage{diagbox}
\usepackage{epsfig}
\usepackage{epsf}
\usepackage{epstopdf}
\usepackage{float}
\usepackage{graphicx}
\usepackage{graphics}
\usepackage{harpoon}
\usepackage{indentfirst}
\usepackage{makecell}
\usepackage{mathrsfs}
\usepackage{mathtools}
\usepackage{multirow}
\usepackage{pifont}
\usepackage{xcolor}

\newcommand{\RM}[1]{\textrm{\uppercase\expandafter{\romannumeral#1}}}

\begin{document}

\title{Relativistic corrections to energy spectrum of hydrogen due to the full one-photon-exchange interaction}

\author{
Zi-Wen Zhang and Hai-Qing Zhou\protect\footnotemark[1]\protect\footnotetext[1]{E-mail: zhouhq@seu.edu.cn} \\
School of Physics, Southeast University, NanJing 211189, China}

\date{\today}

\begin{abstract}
In this work, we present expressions for the full effective potential corresponding to the one-photon exchange interaction within the framework of an effective Schr\"{o}dinger-like equation, which is derived exactly from the Bethe-Salpeter equation in quantum electrodynamics. The final effective potential is expressed in terms of eight scalar functions. When these eight scalar functions are expanded in terms of velocities order by order, we can return to the non-relativistic effective potential systematically organized by velocities. By retaining the exact momentum dependence in the effective potential, we estimate its corrections to the energy spectrum of hydrogen using a highly precise numerical method. The comparison of the numerical results with those obtained using conventional bound-state perturbative theory is discussed. Our calculations suggest that it is possible to accurately account for all relativistic contributions using this method. It would be interesting to extend these calculations to positronium, muonic hydrogen, and scenarios involving nuclear structure and radiative corrections.

\end{abstract}

\maketitle

\section{Introduction}
The study of the energy spectrum of hydrogen-like atoms has been pivotal in the development of quantum mechanics and has remained important for over a century. However, dealing with bound states in a pure quantum field theory, especially when the non-relativistic expansion is not valid, can still be challenging. In the past fifteen years, there have been significant improvements in the precise experimental measurements of the Lamb shifts of hydrogen and muonic hydrogen, presenting many challenges \cite{Pohl-Nature-2010,Antognin-Science-2013,Beyer-Science-2017,
Fleurbaey-PRL-2018,Bezginov-Science-2019,Grinin-Science-2020,Brandt-PRL-2022,Bullis-PRL-2023}. Theoretically, the bound-state perturbative theory is commonly used to estimate energy corrections beyond the Coulomb potential (see recent reviews and books \cite{Eides-Book-2007, Jentschura-Book-2022,Pachucki-RMP-2024} and the references therein). To reliably estimate these corrections, the effective Schrodinger-like equation \cite{Lepage-1978-PRA-Schrodinger-like-equation} or the effective Dirac-like equations \cite{Lepage-1977-PRA-Dirac-like-equation}, which are derived exactly from the Bethe-Salpeter (BS) equation \cite{Salpeter-PR-1951-BS-equation, Gell-Mann-PR-1951-BS-equation} in quantum electrodynamics (QED) or non-relativistic quantum electrodynamics (NRQED) \cite{Lepage-1986-NRQED}, should be employed, as there is no analytical solution for the physical BS equation.

In the bound-state perturbative theory, the interaction kernel in the effective Schr\"{o}dinger-like equation or the effective Dirac-like equations is expanded order by order in terms of the fine structure constant $\alpha_e$, velocities $\vec{p}_i/m_{e,p}$, and $m_e/m_p$, where $\vec{p}_i$ represents the three momenta of the particles in the center-of-mass frame, and $m_e$ and $m_p$ denote the masses of the electron and proton, respectively. The expansion of the interaction kernel in terms of velocities implies that only contributions from low momenta are accurately considered, while contributions from high momenta need to be treated separately.

On the other hand, the wave functions derived from  the Schr\"{o}dinger-like equation with the Coulomb potential decrease  polynomily on the momentum in the momentum space.  These wave functions are used for the entire momentum range in the bound-state perturbative theory. The integration of the wave functions and interaction kernels encompasses contributions from both low and high momenta. This can introduce additional ultraviolet (UV) divergences, which differ somewhat from the usual UV divergences in the scattering amplitudes arising from loop integrals in radiative corrections. To handle these UV divergences, regularization methods are employed \cite{Pachucki-PRA-1997, Czarnecki-PRA-1999, Adkins-PRA-2020}. After considering all the contributions at the same order, these UV divergences eventually cancel each other out, and the final results are independent of the regularization procedure and are free of divergences.

Throughout history, various other methods have also been employed to investigate corrections beyond the Coulomb potential. These include the use of the Dirac equation with an effective potential \cite{Grotch-RMP-1969}, the external field approximation \cite{Eides-Book-2007}, the quasi-potential approach \cite{Faustov-1985-TMP, Faustov-1999-zETF, Martynenko-2005-JETP}, and the Foldy-Wouthuysen transformation method \cite{Foldy-PR-1950, Pachucki-2005-PRA, Zhou-JPB-2023}, among others.

In this work, we focus our discussion within the framework of the effective Schr\"{o}dinger-like equation, which is derived exactly from the Bethe-Salpeter equation in QED \cite{Lepage-1978-PRA-Schrodinger-like-equation}. Different from the usual bound-state perturbative theory applied in hydrogen-like system, we do not expand the interaction kernel in terms of velocities and $m_e/m_p$, but retain them in its  relativistic form. The relativistic form effective potential has correct behavior at high energy or short distance. By taking the one-photon-exchange (OPE) interaction kernel as an example, we express the corresponding relativistic form effective potential by eight scalar functions which includes all relativistic corrections at the order of $\alpha_e$. This effective potential are valid across the entire momentum range and can reproduce the usual low energy behavior. By this potential, the additional UV divergences in the bound-state perturbative theory naturally disappear. Furthermore, we use a high precise numerical method calculate the energy spectrum  by this potential.  This numerical calculation can be regarded as a full result incorporating the OPE interaction, and it provides an interesting basis for comparison with the results obtained using the usual bound-state perturbative theory.

The paper is organized as follows: In Section \RM{2}, we introduce the framework used in our calculations. In Section \RM{3}, we present the numerical results for specific states, showcasing their properties. We also provide a brief discussion on the comparison of these numerical results with those obtained using the bound-state perturbative approach.

\section{Basic Formula}
\subsection{The effective Schr\"{o}dinger-like equation}
\begin{figure}[htbp]
\centering
\includegraphics[width=10.8cm]{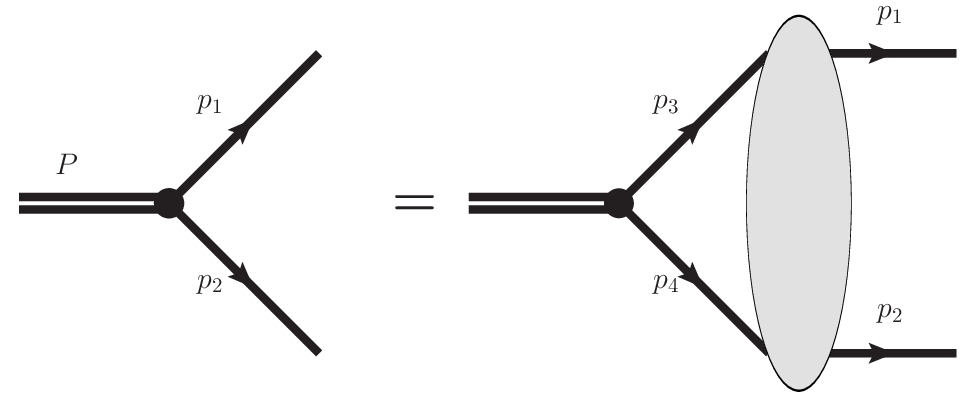}\\
\caption{Diagram for the BS equation.}
\label{Figure-BS-equation}
\end{figure}

In QED, the bound states of the $ep$ system in the center frame can be accurately described by the following effective Schr\"{o}dinger-like equation \cite{Lepage-1978-PRA-Schrodinger-like-equation}:
\begin{equation}
\begin{aligned}
\left[P_0-E^{(e)}_{1}-E^{(p)}_{1}\right] \phi_{\bar{\lambda},\bar{\mu}}(\vec{p}_{1})=\int \frac{d^3 \vec{p}_3}{(2 \pi)^3} i \widetilde{K}_{\bar{\lambda}\lambda,  \bar{\mu}\mu}(\vec{p}_{1}, \vec{p}_{3}, P) \phi_{\lambda,\mu}(\vec{p}_{3}),
\end{aligned}
\label{Eq-Schrodinger-like-equation-1}
\end{equation}
where $P=(P_0,\vec{0})\equiv p_1+p_2=p_3+p_4$,  $p_{1,2,3,4}$ are the momenta of final and intermediate fermions as shown in Fig.\ref{Figure-BS-equation}, $E_{i}^{(e)}\equiv\sqrt{\vec{p}_i^2+m_e^2}$, $E_{i}^{(p)}\equiv\sqrt{\vec{p}_i^2+m_p^2}$, and $\lambda,\mu$ are the indexes related with projected spinors.  The interaction kernel $\widetilde{K}_{\bar{\lambda}\lambda,  \bar{\mu}\mu}(\vec{p}_{1}, \vec{p}_{3}, P)$ is defined as
\begin{equation}
\begin{aligned}
\widetilde{K}_{\bar{\lambda}\lambda,  \bar{\mu}\mu}(\vec{p}_{1}, \vec{p}_{3}, P)&\equiv  \frac{\bar{u}_{\bar{\alpha}}(\vec{p}_{1},m_e,\bar{\lambda}) \bar{u}_{\bar{\beta}}(-\vec{p}_{1},m_p,\bar{\mu})}{\sqrt{4 E^{(e)}_{1} E^{(p)}_{1}}}
\bar{K}_{\bar{\alpha}\alpha,\bar{\beta}\beta}(\vec{p}_{1},\vec{p}_{3}, P) \frac{u_{\alpha}(\vec{p}_{3},m_e,\lambda) u_{\beta}(-\vec{p}_{3}, m_p,\mu)}{\sqrt{4 E^{(e)}_{3}E^{(p)}_{3}}},\\
\end{aligned}
\end{equation}
where $\lambda,\bar{\lambda},\mu,\bar{mu}$ are the helicity of the projected spinors and
\begin{equation}
\begin{aligned}
\bar{K}(\vec{p}_{1}, \vec{p}_{3}, P)&\equiv \bar{K}(p_1, p_3, P)\Big|_{p_{1}^{0}=p_{3}^{0}=\tau_e P_0},\\
\bar{K} &\equiv [I-K_{\textrm{BS}}(G_0-\bar{G}_0)]^{-1} K_{\textrm{BS}},
\end{aligned}
\end{equation}
with $\tau_e\equiv m_e/(m_e+m_p)$ and $K_{\textrm{BS}}$ the usual two-body BS irreducible kernel. Here, we directly make the assumption that the inverse $[I-K_{\textrm{BS}}(G_0-\bar{G}_0)]^{-1}$ exists, and that $\bar{K}$ can be expressed in a perturbative form as follows:
\begin{equation}
\begin{aligned}
\bar{K}(p_1, p_3, P) &=K_{\textrm{BS}}(p_1, p_3, P)+\int \frac{\textrm{d}^4 k}{(2 \pi)^4} K_{\textrm{BS}}(p_1, k, P)[G_0(k,P)-\bar{G}_0(k,P)] K_{\textrm{BS}}(k, p_3, P)+\cdots
\end{aligned}
\end{equation}
where
\begin{equation}
\begin{aligned}
G_0(k, P)&\equiv S^{(e)}_{full}(k,m_e)S^{(p)}_{full}(P-k,m_p),\\
\bar{G}_0(k, P)&\equiv 2 \pi i \delta(k^0)\Big[\frac{\Lambda_{+}^{(e)}(\vec{k}) \Lambda_{+}^{(p)}(-\vec{k})}{P_0-E_{e}(\vec{k})-E_{p}(\vec{k})}-\frac{\Lambda_{-}^{(e)}(\vec{k}) \Lambda_{-}^{(p)}(-\vec{k})}{P_0+E_{e}(\vec{k})+E_{p}(\vec{k})}\Big]\\
&\approx 2 \pi i \delta(k^0)\frac{\Lambda_{+}^{(e)}(\vec{k}) \Lambda_{+}^{(p)}(-\vec{k})}{P_0-E_{e}(\vec{k})-E_{p}(\vec{k})}.
\end{aligned}
\end{equation}
In the above expressions, $S^{(e,p)}_{full}$ are the full propagators of the electron and proton \cite{Jentschura-Book-2022}, respectively, and $\Lambda_{\pm}^{(a)}(\vec{k})$ are
\begin{equation}
\begin{aligned}
\Lambda_{\pm}^{(a)}(\vec{k})&\equiv\left[E^{(a)}(\vec{k}) \gamma^0 \mp(\vec{k} \cdot \vec{\gamma}-m_a)\right]^{(a)} / 2 E^{(a)}(\vec{k}),
\end{aligned}
\end{equation}
with $E^{(a)}(\vec{k})\equiv \sqrt{\vec{k}^2+m_a^2}$.

In the leading order of $\alpha_e$, $\vec{p}_{i}/m_{e,p}$, and $m_e/m_p$, the interaction kernel $\widetilde{K}$ can be chosen as
\begin{equation}
i \widetilde{K}^{(0)}_{\bar{\lambda}\lambda,\bar{\mu}\mu}(\vec{p}_{1}, \vec{p}_{3}, P)\equiv\delta_{\bar{\lambda}\lambda}\delta_{\bar{\mu}\mu}\frac{-e^2 }{|\vec{p}_{1}-\vec{p}_{3}|^2}\frac{1} {[N(\vec{p}_{1}) N(\vec{p}_{3})]^{1 / 2}},
\end{equation}
where
\begin{equation}
\begin{aligned}
N(\vec{p}_{i}) &\equiv\frac{[P_0+E^{(e)}_{i}+E^{(p)}_{i}][P_0^2-(E^{(e)}_{i}+E^{(p)}_{i})^2]}{2 P_0\left[P_0^2-(m_e-m_p)^2\right]}.
\end{aligned}
\end{equation}

By utilizing this effective interaction kernel and expanding the equation in terms of momenta and the bound energy, the effective Schr\"{o}dinger-like equation Eq. (\ref{Eq-Schrodinger-like-equation-1}) simplifies to the Schr\"{o}dinger equation with the Coulomb potential. The contributions from interaction kernels involving higher orders of $\alpha_e$, $\vec{p}_{i}/m_{e,p}$, and $m_e/m_p$ beyond $\widetilde{K}^{(0)}(\vec{p}_{1}, \vec{p}_{3}, P)$ can be estimated systematically using the bound-state perturbative theory, order by order. This approach can also be applied in NRQED, where the power counting and analytical calculations are more straightforward.

\subsection{The full interaction kernel in the leading order of $\alpha_e$}
In the bound-state perturbative theory, the interaction kernels are expanded order by order in terms of $\vec{p}_i/m_{e,p}$, and the corresponding matrix elements often exhibit divergences. To handle these divergences, regular methods are employed during the intermediate calculations. It is important to note that these divergences arise naturally due to the slow decrease of the wave functions of the bound states with respect to momenta $\vec{p}_i$, whereas the expansion of the interaction kernels in terms of momenta is not valid in the high momentum region.

In this study, we adopt a different approach by not expanding the interaction kernel $\widetilde{K}(\vec{p}_{1}, \vec{p}_{3}, P)$ in terms of $\vec{p}_i/m_{e,p}$ and $m_e/m_p$. Instead, we employ a numerical method to directly solve Eq. (\ref{Eq-Schrodinger-like-equation-1}) using the leading-order $\alpha_e$ expression of $\widetilde{K}(\vec{p}_{1}, \vec{p}_{3}, P)$.

In the leading order of $\alpha_e$, only the OPE interaction contributes. As an approximation, we consider the proton to be a point-like particle in this study. Therefore, we have:
\begin{equation}
\begin{aligned}
K^{(\textrm{OPE})}_{\textrm{BS},\bar{\alpha}\alpha,\bar{\beta}\beta}(p_1,p_3,P) &=(-ie\gamma^\mu_{\bar{\alpha}\alpha})D_{\mu\nu}(p_1-p_3)(+ie\gamma^\nu_{\bar{\beta}\beta}),
\end{aligned}
\end{equation}
and
\begin{equation}
\begin{aligned}
\bar{K}^{(\textrm{LO})}(\vec{p}_{1}, \vec{p}_{3}, P) &= K_{\textrm{BS}}^{(\textrm{OPE})}(p_1,p_3, P)\Big|_{p_{1}^{0}=p_{3}^{0}=\tau_e P_0}.
\end{aligned}
\end{equation}
Taking the Coulomb gauge, we have
\begin{equation}
\begin{aligned}
\bar{K}^{(\textrm{LO})}_{\bar{\alpha}\alpha,\bar{\beta}\beta}(\vec{p}_{1},\vec{p}_{3},P)
&=(-ie\gamma^0_{\bar{\alpha}\alpha})\frac{i}{|\vec{q}|^2} (+ie\gamma^0_{\bar{\beta}\beta})+(-ie\gamma^i_{\bar{\alpha}\alpha})[
\frac{iq^iq^j}{|\vec{q}|^4}-\frac{i\delta^{ij}}{{|\vec{q}|^2}}] (+ie\gamma^j_{\bar{\beta}\beta}),
\end{aligned}
\end{equation}
with $q=p_{1}-p_3$. Here, $q^0=0$ due to the special choice of the momenta with $p_{1}^{0}=p_{3}^{0}=\tau_e P_0$.

After performing the necessary calculations, we obtain the following result:
\begin{equation}
\begin{aligned}
i\widetilde{K}_{\bar{\lambda}\lambda, \bar{\mu}\mu}(\vec{p}_{1}, \vec{p}_{3}, P)
&\equiv \chi^{(e)\dag}(\bar{\lambda}) \chi^{(p)\dag}(\bar{\mu}) V(\vec{p}_1,\vec{p}_3) \chi^{(p)}(\mu)\chi^{(e)}(\lambda)\\
&= \sum_{i=1}^{8} \frac{e^2C_i[\chi^{(e)\dag}(\bar{\lambda}) \chi^{(p)\dag}(\bar{\mu}) T_i \chi^{(p)}(\mu)\chi^{(e)}(\lambda)]}{4z_1^2\sqrt{E^{(e)}_{1}E^{(e)}_{3}E^{(p)}_{1}E^{(p)}_{3}X_{1}^{(e)}X_{3}^{(e)}X_{1}^{(p)}X_{3}^{(p)}}},
\end{aligned}
\end{equation}
where $\chi$ is the Pauli spinor and
\begin{equation}
\begin{aligned}
T_1 &= 1, \\
T_2 &= i(\vec{p}_1\times\vec{p}_3)\cdot\sigma^{(e)},\\
T_3 &= i(\vec{p}_1\times\vec{p}_3)\cdot\sigma^{(p)},\\
T_4 &= \vec{p}_1\cdot\sigma^{(e)}\vec{p}_1\cdot\sigma^{(p)},\\
T_5 &= \vec{p}_1\cdot\sigma^{(e)}\vec{p}_3\cdot\sigma^{(p)},\\
T_6 &= \vec{p}_3\cdot\sigma^{(e)}\vec{p}_1\cdot\sigma^{(p)},\\
T_7 &= \vec{p}_3\cdot\sigma^{(e)}\vec{p}_3\cdot\sigma^{(p)},\\
T_8 &= \sigma^{(e)}\cdot\sigma^{(p)},
\end{aligned}
\end{equation}
and
\begin{equation}
\begin{aligned}
C_1 &=(z_2-z_1)(\vec{p}_1\cdot \vec{p}_3)^2-z_1z_3\vec{p}_1\cdot \vec{p}_3-z_2|\vec{p}_1|^2|\vec{p}_3|^2-z_1X_{1}^{(e)}X_{3}^{(e)}X_{1}^{(p)}X_{3}^{(p)}, \\
C_2&=-(E^{(e)}_1+E^{(e)}_3) (X^{(p)}_1+X^{(p)}_3) \vec{p}_1\cdot \vec{p}_3
+z_1 \vec{p}_1\cdot \vec{p}_3+(m_e z_1+t_e )(X^{(p)}_1+X^{(p)}_3)+z_1X^{(p)}_1 X^{(p)}_3 ,\\
C_3&=-(E^{(p)}_1+E^{(p)}_3) (X^{(e)}_1+X^{(e)}_3) \vec{p}_1\cdot \vec{p}_3
+z_1 \vec{p}_1\cdot \vec{p}_3+(m_p z_1+t_p )(X^{(e)}_1+X^{(e)}_3)+z_1X^{(e)}_1 X^{(e)}_3 ,\\
C_4 &= 2E^{(e)}_{3}E^{(p)}_{3}\vec{p}_1\cdot \vec{p}_3+E^{(e)}_{1}(E^{(p)}_{1}-E^{(p)}_{3})|\vec{p}_3|^2-E^{(e)}_{3}(m_pz_1+t_p)-m_eX_{3}^{(p)}z_1 ,\\
C_5 &= -\vec{p}_1\cdot \vec{p}_3[E^{(e)}_{1}(E^{(p)}_{1}+E^{(p)}_{3})+E^{(e)}_{3}(E^{(p)}_{3}-E^{(p)}_{1})]+E^{(e)}_{1}E^{(p)}_{3}z_4+E^{(e)}_{1}m_pz_1+m_eX_{3}^{(p)}z_1 ,\\
C_6 &= -\vec{p}_1\cdot \vec{p}_3[E^{(e)}_{1}(E^{(p)}_{1}-E^{(p)}_{3})+E^{(e)}_{3}(E^{(p)}_{3}+E^{(p)}_{1})]+E^{(e)}_{3}E^{(p)}_{1}z_4+E^{(e)}_{3}m_pz_1+m_eX_{1}^{(p)}z_1 ,\\
C_7 &= 2E^{(e)}_{1}E^{(p)}_{1}\vec{p}_1\cdot \vec{p}_3+E^{(e)}_{3}(E^{(p)}_{3}-E^{(p)}_{1})|\vec{p}_1|^2-E^{(e)}_{1}(m_pz_1+t_p)-m_eX_{1}^{(p)}z_1 ,\\
C_8 &=(-\vec{p}_1\cdot \vec{p}_3(E^{(e)}_{1}+E^{(e)}_{3})+m_ez_1+t_e)(-\vec{p}_1\cdot \vec{p}_3(E^{(p)}_{1}+E^{(p)}_{3})+m_pz_1+t_p),
\end{aligned}
\end{equation}
with
\begin{equation}
\begin{aligned}
X_{i}^{(a)}&\equiv E_{i}^{(a)}+m_a,\\
z_1&\equiv |\vec{p}_1|^2+|\vec{p}_3|^2-2\vec{p}_1\cdot\vec{p}_3,\\
z_2&\equiv (X_{1}^{(e)}+X_{3}^{(e)})(X_{1}^{(p)}+X_{3}^{(p)}),\\
z_3&\equiv X_{1}^{(e)}X_{3}^{(e)}+X_{1}^{(p)}X_{3}^{(p)},\\
z_4&\equiv |\vec{p}_1|^2+|\vec{p}_3|^2,\\
t_e&\equiv E^{(e)}_{1}|\vec{p}_3|^2+E^{(e)}_{3}|\vec{p}_1|^2,\\
t_p&\equiv E^{(p)}_{1}|\vec{p}_3|^2+E^{(p)}_{3}|\vec{p}_1|^2.
\end{aligned}
\end{equation}

By expanding the potential $V$ in terms of $\vec{p}_{i}/m_{e,p}$, it can be verified that the leading-order term corresponds to the Coulomb potential, while the next-leading-order terms correspond to the Breit potential. The higher-order terms correspond to the effective potential beyond the Breit potential. Similarly, the expansion results also correspond to the amplitude in NRQED at the leading order of $\alpha_e$, with higher orders of $\vec{p}_i/m_{e,p}$.

\subsection{Energy correction in the bound-state perturbative theory}

Expanding $H_p \equiv V + (E_1^{(e)}+E_1^{(p)}-m_e-m_p)(2\pi)^3\delta^3(\vec{p}_1-\vec{p}_3)$ in terms of momenta yields the following results:
\begin{equation}
\begin{aligned}
H_p&=H_p^{(2)}+H_p^{(4)}+....,
\end{aligned}
\end{equation}
where
\begin{equation}
\begin{aligned}
H_p^{(2)}&=[\frac{\vec{p}_1^2}{2m_e}+\frac{\vec{p}_1^2}{2m_p}](2\pi)^3\delta^3(\vec{p}_1-\vec{p}_3)-\frac{e^2}{\vec{q}^2},\\
H_p^{(4)}&=H^{(4)}_{p,0}+H^{(4)}_{p,1}+H^{(4)}_{p,2},
\end{aligned}
\end{equation}
with
\begin{equation}
\begin{aligned}
H^{(4)}_{p,0}(\vec{p}_1,\vec{q})=&-\Big[\frac{\vec{p}_1^4}{8m_e^3}+\frac{\vec{p}_1^4}{8m_p^3}\Big](2\pi)^3\delta^3(\vec{p}_1-\vec{p}_3)
+\frac{e^2}{8}\Big[\frac{1}{m_e^2}+\frac{1}{m_p^2}\Big]
+\frac{e^2}{m_em_p}\Big[\frac{(\vec{q}\cdot\vec{p}_1)^2}{\vec{q}^4}-\frac{\vec{p}_1^2}{\vec{q}^2}\Big], \\
H^{(4)}_{p,1}(\vec{p}_1,\vec{q})=&\Big[\frac{e^2}{4m_e^2}+\frac{e^2}{2m_em_p}\Big]\frac{1}{\vec{q}^2}i\sigma^{(e)}\cdot(\vec{q}\times\vec{p}_1),\\
H^{(4)}_{p,2}(\vec{p}_1,\vec{q})=&-\frac{e^2}{4m_em_p}\Big[\frac{\sigma^{(e)}\cdot\vec{q}\sigma^{(p)}\cdot\vec{q}}{\vec{q}^2}-\sigma^{(e)}\cdot\sigma^{(p)}\Big]
+\Big[\frac{e^2}{4m_p^2}+\frac{e^2}{2m_em_p}\Big]\frac{i\sigma^{(p)}\cdot(\vec{q}\times\vec{p}_1)}{\vec{q}^2}.
\end{aligned}
\end{equation}
Here, $H_p^{(4)}$ corresponds to the Breit potential in momentum space.

In the perturbative bound-state theory, the energy contribution of the hydrogen due to the OPE interaction can be written as:
\begin{equation}
\begin{aligned}
E_n=E_n^{(2)}+E_n^{(4)}+ E_n^{(6)}+.....
\end{aligned}
\end{equation}
where
\begin{equation}
\begin{aligned}
E_n^{(2)}&=\langle n,l,j,F|H^{(2)}| n,l,j,F \rangle=-\frac{\alpha_e^2\mu}{2n^2},\\
E_n^{(4)}&= \langle n,l,j,F |H^{(4)}| n,l,j,F \rangle, \\
E_n^{(6)}&= \langle n,l,j,F |H^{(4)} Q(E_n^{(2)}-H^{(2)})^{-1} Q H^{(4)}| n,l,j,F \rangle+\langle n,l,j,F |H^{(6)}| n,l,j,F \rangle.
\end{aligned}
\end{equation}
Here, $\alpha_e\equiv \frac{e^2}{4\pi},\mu=\frac{m_em_p}{m_e+m_p}$ and $Q$ is a projection operator on a subspace orthogonal to $|n,l,j,F \rangle$.

The calculation of $E_{n,i}^{(4)}$ can be performed directly, and the analytic results are expressed as:
\begin{equation}
\begin{aligned}
E_n^{(4)} & = E_{n,0}^{(4)}+E_{n,1}^{(4)}+E_{n,2}^{(4)},
\end{aligned}
\end{equation}
where
\begin{equation}
\begin{aligned}
E_{n,0}^{(4)} &= \frac{\alpha_e^4\mu}{2n^3}\Big[1-\frac{(2-6n)\mu}{n(m_e+m_p)}\Big],\\
E_{n,1}^{(4)} &= \frac{2\alpha_e^4\mu m_p(2m_e+m_p)}{n^3(m_e+m_p)^2} \frac{j-l}{(2 l+1)(2j+1)}(1-\delta_{0 l}),\\
E_{n,2}^{(4)} &=
\frac{\alpha_e^4\mu^2}{n^3(m_e+m_p)}\frac{1}{(2l+1)(2F+1)}\begin{cases}
\frac{4}{3}[F^2+F-\frac{3}{2}]\delta_{0l}, &l=0 \\
-\frac{1}{l}, &l\neq0, F=l+1 \\
-\frac{1}{(l+1)}, &l\neq0, F=l-1.
\end{cases}
\end{aligned}
\end{equation}

The calculation of $E_n^{(6)}$ is somewhat complex \cite{Zhong-PRA-2018,Zhou-PRA-2019,Haidar-PRA-2020,Adkins-PRL-2023}, and there is a UV divergence in the middle matrix elements for the $S$ wave, as explained above. In this work, we will not compare these results.

\subsection{The form of the wave function}
The wave function in Eq. (\ref{Eq-Schrodinger-like-equation-1}) can be written as:
\begin{equation}
\begin{aligned}
\phi_{\bar{\lambda},\bar{\mu}}(\vec{p}_1)=\sum_{\bar{\alpha},\bar{\beta}}
\chi^{(e)\dag}_{\bar{\alpha}}(\bar{\lambda})\chi^{(p)\dag}_{\bar{\beta}}(\bar{\mu})\Phi_{\bar{\alpha},\bar{\beta}}(\vec{p}_1),
\end{aligned}
\end{equation}
where $\chi$ is the Pauli spinor and $\bar{\lambda},\bar{\mu}$ are the spin of the Pauli spinors. This results in the following equation:
\begin{equation}
\begin{aligned}
\left[P_0-E^{(e)}_{1}-E^{(p)}_{1}\right] \Phi_{\bar{\alpha},\bar{\beta}}(\vec{p}_1)&=\int \frac{d^3 \vec{p}_3}{(2 \pi)^3} V_{\bar{\alpha}\alpha,\bar{\beta}\beta}(\vec{p}_1,\vec{p}_3)\Phi_{\alpha,\beta}(\vec{p}_3).
\end{aligned}
\label{Eq-Schrodinger-like-equation-2}
\end{equation}

To compare with the results in the references, we label the state of the system as $nL_{j}^{F}$. For $F=j\pm\frac{1}{2}=l\pm1$, we have the following form for $\Phi_{\alpha,\beta}(\vec{p}_3)$:
\begin{equation}
\begin{aligned}
\Phi_{\alpha,\beta}(\vec{p}_3) = \sum_{s_z^e,s_z^p}\chi_{\alpha}(s_z^{p})\chi_{\beta}(s_z^{e})\langle jj_z|lm,ss_z^{e}\rangle \langle FF_z|jj_z,ss_z^{p}\rangle \Phi_{nlm}(\vec{p}_3).
\label{Eq-wave-function-1}
\end{aligned}
\end{equation}
When the effective potential $V$ does not include spin-dependent terms, the above form of the wave function gives the same result as the following form:
\begin{equation}
\begin{aligned}
\Phi_{\alpha,\beta}(\vec{p}_3) \rightarrow \Phi_{nlm}(\vec{p}_3).
\label{Eq-wave-function-3}
\end{aligned}
\end{equation}

We would also like to mention that the above form of the wave functions corresponds to the quantum numbers in non-relativistic quantum mechanics. When discussing states in QED, one can consider $J^P$ as good quantum numbers to determine the most general form of the wave function $\Phi_{\alpha,\beta}$ for solving the equation. In this case, effects such as the mixing of $S$ wave and $D$ wave naturally appear. However, in this work, we limit our discussion to the relativistic effects when considering the above form of the wave functions.

\subsection{Numerical method}

To calculate the energy contributions beyond the bound-state perturbative theory, we expand $\Phi_{nlm}(\vec{p})$ as:

\begin{equation}
\begin{aligned}
\Phi_{nlm}(\vec{p})\approx \sum_{i=l+1}^{n_{max}}c_{ni} \phi^{(0)}_{ilm}(\vec{p})\equiv \sum_{i=l+1}^{n_{max}}c_{ni} \phi^{(0)}_{il}(|\vec{p}|)Y_{lm}(\Omega_{\vec{p}}),
\end{aligned}
\end{equation}
where
\begin{equation}
\begin{aligned}
\phi^{(0)}_{il}(|\vec{p}|)\equiv \bar{p}^l a^{3 / 2} N_{i l} \frac{l!}{\sqrt{2 \pi}} \times \frac{4^{l+1}}{(\bar{p}^2+\delta^2)^{l+2}} C_{n-l-1}^{l+1}(\frac{\bar{p}^2-\delta^2}{\bar{p}^2+\delta^2}),
\end{aligned}
\end{equation}
and
\begin{equation}
\quad N_{i l}=\frac{2}{i^2} \sqrt{\frac{(i-l-1) !}{(i+l) !}}, \quad \bar{p}=a p,
\end{equation}
with  $C$ the Gegenbauer polynomial and $a=\frac{1}{\mu e^2}$ the Bohr radius.

By choosing a specific value of $n_{max}$, one can calculate the following matrix:
\begin{equation}
\begin{aligned}
\int d^3\vec{p}_1\Phi^{(0)\dag}_{\bar{\alpha},\bar{\beta};\bar{i}}(\vec{p}_1)[E^{(e)}_{1}+E^{(p)}_{1}]
\delta_{\bar{\alpha}\alpha}\delta_{\bar{\beta}\beta}\Phi^{(0)}_{\alpha,\beta;i}(\vec{p}_1)+
\int \frac{d^3\vec{p}_1d^3\vec{p}_3}{(2\pi)^3} \Phi^{(0)\dag}_{\bar{\alpha},\bar{\beta};\bar{i}}(\vec{p}_1)
V_{\bar{\alpha}\alpha,\bar{\beta}\beta}(\vec{p}_1,\vec{p_3})\Phi^{(0)}_{\alpha,\beta;i}(\vec{p}_3),
\end{aligned}
\end{equation}
where $\Phi^{(0)}_{\alpha,\beta;i}(\vec{p}_1)$ is simply $\Phi_{\alpha,\beta}(\vec{p}_1)$ in Eq. (\ref{Eq-wave-function-1}) after replacing $\Phi_{nlm}(\vec{p}_1)$ with $\phi^{(0)}_{il}(|\vec{p}_1|)Y_{lm}$. The integration over the angles $\Omega_{\vec{p}1}$ and $\Omega_{\vec{p}_3}$ can be performed analytically, while the integration over $|\vec{p_1}|$ and $|\vec{p}_3|$ can be done numerically with high precision. After diagonalizing this matrix, the energy spectrum $E_{n}\equiv P_{0,n}-m_e-m_p$ is obtained approximately.

We would like to highlight a significant difference between the method described above and the bound state perturbative theory. In the latter, non-relativistic potentials are utilized order by order. These non-relativistic potentials can lead to additional UV divergences in certain matrix elements, which are only cancelled when all intermediate states are considered. In contrast, our calculation correctly accounts for the higher energy behavior of the effective potential, and there are no additional UV divergences; a finite $n_{max}$  can yield precise results. We will elaborate on this property in detail in the next section.

\section{Numerical Results and discussion}
In our numerical calculation, we set $n_{max}=100$, and the physical constants are taken as $m_e=0.510998950$ MeV, $m_p=938.27208816$ MeV, and $1/\alpha_e=137.035999084$. The relative precision of the numerical calculation for each matrix element reaches $10^{-20}$, ensuring that the absolute precision of the matrix elements is better than $10^{-18}$ eV. This guarantees the reliability of the numerical results. The numerical precision is also tested for the matrix elements involving the Coulomb and Breit potentials.

To compare the results with those obtained using bound-state perturbative theory, we decompose the effective potential into three terms as follows:
\begin{equation}
\begin{aligned}
V&\equiv V_{0}+V_{1}+V_{2},
\end{aligned}
\end{equation}
where $V_0$ is spin-independent, $V_1$ depends only on the electron spin, and $V_2$ depends on the proton spin. We label the corresponding energy contributions due to $V_0,V_{0}+V_{1}, V_{0}+V_{1}+V_{2}$ as  $E_{n,0},E_{n,1},E_{n,2}$, respectively.
The contributions beyond the Breit potential are expressed by the following quantities:
\begin{equation}
\begin{aligned}
\Delta E_{n,0}&\equiv E_{n,0}-E_{n}^{(2)}-E_{n,0}^{(4)},\\
\Delta E_{n,1}&\equiv E_{n,1}-E_{n}^{(2)}-E_{n,0}^{(4)}-E_{n,1}^{(4)},\\
\Delta E_{n,2}&\equiv E_{n,2}-E_{n}^{(2)}-E_{n,0}^{(4)}-E_{n,1}^{(4)}-E_{n,2}^{(4)}.
\end{aligned}
\end{equation}

\begin{table}[htbp]
\renewcommand\arraystretch{1}
\centering
\begin{tabular}{p{2.0cm}<{\centering}p{4.0cm}<{\centering}p{4.0cm}<{\centering}p{4.0cm}<{\centering}}
\hline
\hline
\diagbox[width=2cm,height=1.1cm]{$n$}{peV}&$\Delta E_{n,0}(S)$ & $\Delta E_{n,0}(P)$ &  $\Delta E_{n,0}(D)$  \\
\hline
$n=1$ &-2349361 &  - & -\\
\hline
$n=2$ &-289552 & 813  &  - \\
\hline
$n=3$ &-85154 & 424  & 13\\
\hline
$n=4$ &-35738  &242  & 16 \\
\hline
$n=5$ &-18225  & 151  &14\\
\hline
\hline
\end{tabular}
\caption{The numeric results for the energy corrections $\Delta E_{n,0}(l)$, where the unit is peV ($10^{-12}$eV).}
\label{Tab-DeltaE0}
\end{table}

\begin{table}[htbp]
\renewcommand\arraystretch{1}
\centering
\begin{tabular}{p{2.0cm}<{\centering}p{4.0cm}<{\centering}p{4.0cm}<{\centering}p{4.0cm}<{\centering}}
\hline
\hline
\diagbox[width=2cm,height=1.1cm]{$n$}{peV}&$\Delta E_{n,1}(S)$ & $\Delta E_{n,1}(P)$ &  $\Delta E_{n,1}(D)$  \\
\hline
$n=1$ &-2349361  &  - & -\\
\hline
$n=2$ &-289552  & -44  &  - \\
\hline
$n=3$ &-85154  &  50   & 3\\
\hline
$n=4$ &-35738  &  48  &  6\\
\hline
$n=5$ &-18225 & 37    &  6 \\
\hline
\hline
\end{tabular}
\caption{The numeric results for the energy correction $\Delta E_{n,1}(l)$ with $j=l+1/2$, where the unit is peV ($10^{-12}$eV).}
\label{Tab-DeltaE1}
\end{table}

\begin{table}[htbp]
\renewcommand\arraystretch{1}
\centering
\begin{tabular}{p{2.0cm}<{\centering}p{4.0cm}<{\centering}p{4.0cm}<{\centering}p{4.0cm}<{\centering}}
\hline
\hline
\diagbox[width=2cm,height=1.1cm]{$n$}{peV}&$\Delta E_{n,2}(S)$ & $\Delta E_{n,2}(P)$ &  $\Delta E_{n,2}(D)$  \\
\hline
$n=1$ &-2357360 &  - & -\\
\hline
$n=2$ &-290555 & -45  &  - \\
\hline
$n=3$ &-85452 & 50   & 3\\
\hline
$n=4$ &-35864 & 48  &  6 \\
\hline
$n=5$ &-18289 & 37   &  6 \\
\hline
\hline
\end{tabular}
\caption{The numeric results for the energy correction $\Delta E_{n,2}(l)$ with $F=j+1/2=l+1$,where the unit is peV ($10^{-12}$eV).}
\label{Tab-DeltaE2}
\end{table}

We present the numeric results for $\Delta E_{n,i}$ in Tabs. \ref{Tab-DeltaE0}, \ref{Tab-DeltaE1} and \ref{Tab-DeltaE2}, respectively, where $l=0,1,2$ or $S,P,D$ waves, $j=l+1/2$ and $F=l+1$ are considered. For the purpose of a more direct comparison, we also provide the contributions with specific orders as follows:
\begin{equation}
\begin{aligned}
\alpha_e^5\mu  &\sim  11~\mu\textrm{eV},\\
\alpha_e^6\mu &\sim  77~\textrm{neV} ,\\
\alpha_e^6\mu \frac{m_e}{m_p} &\sim 42~\textrm{peV}.
\end{aligned}
\end{equation}

The results in Tab. \ref{Tab-DeltaE0}, \ref{Tab-DeltaE1} and \ref{Tab-DeltaE2} clearly demonstrate a notable property whereby the contributions in the $S$ wave are significantly larger than those in the $P$ and $D$ waves.

In terms of magnitude, the contributions $\Delta E_{n,0}(S)$ are approximately $-\frac{\alpha_e^5\mu}{4.5n^3}\sim -\frac{30\alpha_e^6\mu}{n^3}$, which are larger than the contributions at the order of $\frac{\alpha_e^6\mu}{n^3}$. The contributions $\Delta E_{n,0}(P)$ are on the order of $\frac{\alpha_e^6\mu}{20n^2}$. On the other hand, the contributions $\Delta E_{n,0}(D)$ weakly depend on $n$ and are much smaller than $\alpha_e^6\mu$.

The results for $\Delta E_{n,1}(S)$ and $\Delta E_{n,2}(S)$ are similar to $\Delta E_{n,0}(S)$, indicating that the contributions from spin-independent terms are most significant for the $S$ wave. This is expected, as the contributions from electron spin are zero for the $S$ wave, and the contributions from proton spin are greatly suppressed.

The results for $\Delta E_{n,1}(P)$ and $\Delta E_{n,2}(P)$ are similar to each other and significantly differ from $\Delta E_{n,0}(P)$. This indicates that in the $P$ wave, the contributions from the electron spin are of the same order as the spin-independent contributions. The spin-dependent contributions in the $D$ wave are similar to those in the $P$ wave.

To illustrate the contributions more explicitly, we define the following terms:
\begin{equation}
\begin{aligned}
\Delta E_{n}^{fin}&\equiv \Delta E_{n,1}-\Delta E_{n,0} = E_{n,1}-E_{n,0}-E_{n,1}^{(4)} ,\\
\Delta E_{n}^{hyf}&\equiv \Delta E_{n,2}-\Delta E_{n,1} = E_{n,2}-E_{n,1}-E_{n,2}^{(4)}.
\end{aligned}
\end{equation}
These terms reflect the corrections to the fine structure and hyperfine structure beyond the Breit potential, respectively.

\begin{table}[htbp]
\renewcommand\arraystretch{1}
\centering
\begin{tabular}{p{2.0cm}<{\centering}p{4.0cm}<{\centering}p{4.0cm}<{\centering}p{4.0cm}<{\centering}}
\hline
\hline
\diagbox[width=2cm,height=1.1cm]{$n$}{peV}&$\Delta E_{n}^{fin}(S)$ & $\Delta E_{n}^{fin}(P)$ &  $\Delta E_{n}^{fin}(D)$  \\
\hline
$n=1$ &0 & - & -\\
\hline
$n=2$ &0 & -857   &  - \\
\hline
$n=3$ &0 & -374    & -10  \\
\hline
$n=4$ &0  & -194  & -10   \\
\hline
$n=5$ &0  & -114   & -8 \\
\hline
\hline
\end{tabular}
\caption{Numeric results for the energy correction $\Delta E_{n}^{fin}(l)$ with $j=l+1/2$, where the unit is peV ($10^{-12}$eV).}
\label{Tab-DeltaE-fin}
\end{table}

\begin{table}[htbp]
\renewcommand\arraystretch{1}
\centering
\begin{tabular}{p{2.0cm}<{\centering}p{4.0cm}<{\centering}p{4.0cm}<{\centering}p{4.0cm}<{\centering}}
\hline
\hline
\diagbox[width=2cm,height=1.1cm]{$n$}{peV}&$\Delta E_{n}^{hyf}(S)$ & $\Delta E_{n}^{hyf}(P)$ &  $\Delta E_{n}^{hyf}(D)$  \\
\hline
$n=1$ &-7999  &  - & -\\
\hline
$n=2$ &-1003  & $-0.66$  &  - \\
\hline
$n=3$ &-298  &  $-0.27$   & $-0.007$ \\
\hline
$n=4$ &-126   & $-0.14$   & $-0.007$   \\
\hline
$n=5$ &-65  &  $-0.08$  & $-0.005$  \\
\hline
\hline
\end{tabular}
\caption{Numeric results for the energy correction $\Delta E_{n}^{hyf}(l)$ with $F=l+1$, where the unit is peV ($10^{-12}$eV).}
\label{Tab-DeltaE-hyf}
\end{table}

The numeric results for $\Delta E_{n}^{fin}$ and $\Delta E_{n}^{hyf}$ are presented in Tab. \ref{Tab-DeltaE-fin} and  \ref{Tab-DeltaE-hyf}, respectively. The numerical results in Tab. \ref{Tab-DeltaE-fin} indicate that the contributions to the fine structure beyond the Breit potential are approximately $-\frac{\alpha_e^6\mu}{22n^2}$ in the $P$ wave and around $-10$ peV in the $D$ wave. These values are much smaller compared to the contributions of $-\frac{\alpha_e^6\mu}{n^2}$. The numeric results in Tab. \ref{Tab-DeltaE-hyf} show that the contributions to the hyperfine structure beyond the Breit potential are approximately $-\frac{200\alpha_e^6\mu}{n^3}\frac{m_e}{m_p}$ for the $S$ wave, while the contributions are negligible for the $P$ wave and $D$ wave cases.

\begin{figure}[htbp]
\centering
\includegraphics[width=10.8cm]{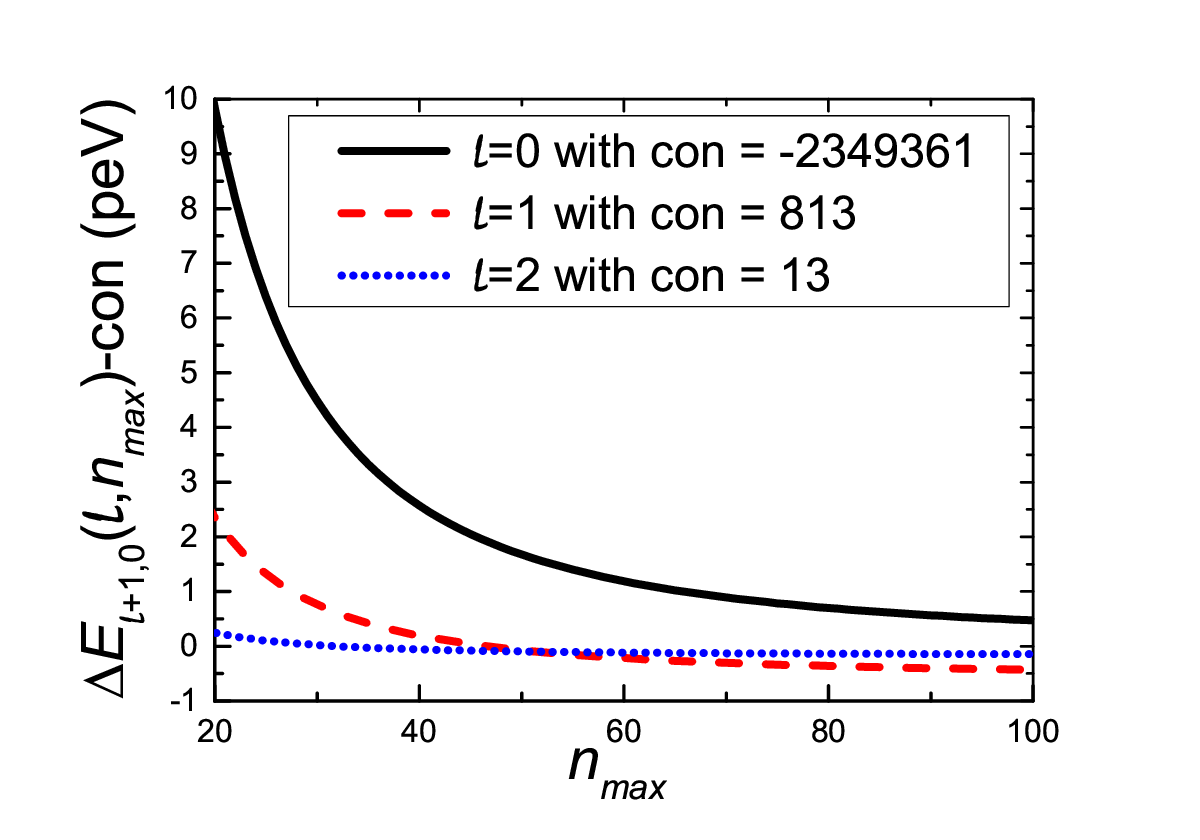}\\
\caption{Numeric results for $\Delta E_{l+1,0}(l,n_{\text{max}})$ vs. $n_{max}$ with $l=0,1$ and $2$, where the unit is peV ($10^{-12}$eV).}
\label{Figure-DeletaE0-vs-nmax}
\end{figure}

To demonstrate the uncertainty arising from the approximation of finite $n_{\text{max}}$, we plot $\Delta E_{l+1,0}(l,n_{max})$ as a function of $n_{max}$ in Fig. \ref{Figure-DeletaE0-vs-nmax}. The results clearly indicate that the uncertainty is smaller than $0.5$peV when $n_{max}>60$, suggesting that the approximation is reliable.

In our calculation, we do not expand the OPE interaction kernel order by order in momenta, but instead solve the effective Schr\"{o}dinger-like equation using numerical method. This approach enables us to include all contributions from ladder diagrams with the approximated propagator $\bar{G}_0$ and the full photon propagator, while excluding the crossed diagrams. Consequently, the comparison of our results with those obtained from bound-state perturbation theory is not straightforward.

For a consistent comparison between the two approaches, one would need to separate the contributions in bound-state perturbation theory based on the diagram types and momentum regions, and then compare them with our results to assess the relativistic contributions obtained through the expansion in bound-state perturbation theory. However, such a direct separation and comparison fall beyond the scope of this work.

In summary, the effective potential associated with the full OPE interaction in momentum space is expressed through eight scalar functions. The expansions of these scalar functions directly correspond to the $ep\rightarrow ep$ amplitude in NRQED or the quasi-potential at order $\alpha_e$ and any desired order of momenta. Our precise numerical calculations suggest that it is possible to capture all relativistic contributions using this method. Extending these calculations to positronium, muonic hydrogen, and cases involving nuclear structure and radiative corrections would be interesting directions for future research.

\section{Acknowledgments}
H.Q.Z. would like to thank Zhi-Hui Guo for his helpful suggestions and discussions. This work is funded in part by the National Natural Science Foundations of China under Grants No. 12075058, No. 12150013.


\begin{thebibliography}{99}

\bibitem {Pohl-Nature-2010}
Pohl R, {\it et al.}, Nature  {\bf 466}, 213 (2010).


\bibitem{Antognin-Science-2013}
Aldo Antognini, {\it et al.}, Science {\bf 339}, 417 (2013).

\bibitem{Beyer-Science-2017}
A. Beyer, L. Maisenbacher, A. Matveev, R. Pohl, K. Khabarova, A. Grinin, T. Lamour,
D.C. Yost, T.W. Hansch, N. Kolachevsky et al., Science 358, 79 (2017).

\bibitem{Fleurbaey-PRL-2018}
H.~Fleurbaey, S.~Galtier, S.~Thomas, M.~Bonnaud, L.~Julien, F.~m.~c.~Biraben, F~.~m.~c.~Nez,
M.~Abgrall, J.~Guena, Phys. Rev. Lett. 120, 183001 (2018).

\bibitem{Bezginov-Science-2019}
N. Bezginov, T. Valdez, M. Horbatsch, A. Marsman, A.C. Vutha, E.A. Hessels, Science
365, 1007 (2019).


\bibitem{Grinin-Science-2020}
A. Grinin, A. Matveev, D.C. Yost, L. Maisenbacher, V. Wirthl, R. Pohl, T.W. Hansch,
T. Udem, Science 370, 1061 (2020).



\bibitem{Brandt-PRL-2022}
A.~D.~Brandt, S.~F.~Cooper, C.~Rasor, Z.~Burkley, D.~C.~Yost and A.~Matveev,
Phys. Rev. Lett. \textbf{128}, 023001 (2022).

\bibitem{Bullis-PRL-2023}
R.~G.~Bullis, C.~Rasor, W.~L.~Tavis, S.~A.~Johnson, M.~R.~Weiss and D.~C.~Yost,
Phys. Rev. Lett. \textbf{130}, 203001 (2023).




\bibitem{Eides-Book-2007}
M.~I.~Eides, H.~Grotch and V.~A.~Shelyuto,
``Theory of Light Hydrogenic Bound States,''
Springer Tracts Mod. Phys. \textbf{222},1 (2007),
ISBN 978-3-540-45269-0, 978-3-540-45270-6.


\bibitem{Jentschura-Book-2022}
U.~D.~Jentschura and G.~S.~Adkins,
``Quantum Electrodynamics: Atoms, Lasers and Gravity,''
World Scientific, 2022,
ISBN 978-981-12-5225-9, 978-981-12-5227-3.


\bibitem{Pachucki-RMP-2024}
K.~Pachucki, V.~Lensky, F.~Hagelstein, S.~S.~Li Muli, S.~Bacca and R.~Pohl,
Rev. Mod. Phys. \textbf{96}, 015001 (2024).





\bibitem{Lepage-1978-PRA-Schrodinger-like-equation}
H. W.~E.~Caswell and G.~P.~Lepage,
Phys. Rev. A \textbf{18}, 810 (1978).


\bibitem{Lepage-1977-PRA-Dirac-like-equation}
G.~P.~Lepage,
Phys. Rev. A \textbf{16}, 863 (1977).

\bibitem{Salpeter-PR-1951-BS-equation}
E.~E.~Salpeter and H.~A.~Bethe, Phys. Rev. \textbf{84}, 1232 (1951).

\bibitem{Gell-Mann-PR-1951-BS-equation}
M. Gell-Mann and F. Low, Phys. Rev. \textbf{84}, 350 (1951).

\bibitem{Lepage-1986-NRQED}
W. Caswell and G. Lepage, Phys. Lett. B 167, 437
(1986).




\bibitem{Pachucki-PRA-1997}
K.~Pachucki,
Phys. Rev. A \textbf{56}, 297 (1997).



\bibitem{Czarnecki-PRA-1999}
A.~Czarnecki, K.~Melnikov and A.~Yelkhovsky,
Phys. Rev. A \textbf{59}, 4316 (1999).

\bibitem{Adkins-PRA-2020}
G.~S.~Adkins, M.~F.~Alam, C.~Larison and R.~Sun,
Phys. Rev. A \textbf{101}, 042511 (2020).


\bibitem{Grotch-RMP-1969}
H.~Grotch and D.~r.~Yennie,
Rev. Mod. Phys. \textbf{41}, 350 (1969).



\bibitem{Faustov-1985-TMP}
R.~N.~Faustov, A.~P.~Martynenko, Teor.Mat.Fiz. 64,179 (1985).

\bibitem{Faustov-1999-zETF}
 R.~N.~Faustov, A.~P.~Martynenko, Zh.Eksp.Teor.Fiz. 115, 1221 (1999).

\bibitem{Martynenko-2005-JETP}
A.~P.~Martynenko, Jour. Exp. Theor. Phys. 101, 1021 (2005).


\bibitem{Foldy-PR-1950}
L. L. Foldy and S. A. Wouthuysen, Phys. Rev. 78, 29 (1950).


\bibitem{Pachucki-2005-PRA}
K.~Pachucki, Phys. Rev. A \textbf{71}, 012503 (2005).

\bibitem{Zhou-JPB-2023}
W.~Zhou, X.~Mei and H.~Qiao,
J. Phys. B \textbf{56}, 045001 (2023).





\bibitem{Zhong-PRA-2018}
Z.~X.~Zhong, W.~P.~Zhou and X.~S.~Mei,
Phys. Rev. A \textbf{98}, no.3, 032502 (2018)
[erratum: Phys. Rev. A \textbf{101}, no.6, 069901 (2020)].


\bibitem{Zhou-PRA-2019}
W.~Zhou, X.~Mei and H.~Qiao,
Phys. Rev. A \textbf{100}, 012513 (2019).


\bibitem{Haidar-PRA-2020}
M.~Haidar, Z.~X.~Zhong, V.~I.~Korobov and J.~P.~Karr,
Phys. Rev. A \textbf{101}, 022501 (2020).


\bibitem{Adkins-PRL-2023}
G.~S.~Adkins, J.~Gomprecht, Y.~Li and E.~Shinn,
Phys. Rev. Lett. \textbf{130}, 023004 (2023).



\end{thebibliography}
\end{document}